\documentclass[prl,twocolumn,aps,amsmath,amssymb]{revtex4}

\usepackage{graphicx}
\usepackage{dcolumn}
\usepackage{bm}
\usepackage{amsmath}
\usepackage{amsfonts}
\newcommand{\drawsquare}[2]{\hbox{%
\rule{#2pt}{#1pt}\hskip-#2pt
\rule{#1pt}{#2pt}\hskip-#1pt
\rule[#1pt]{#1pt}{#2pt}}\rule[#1pt]{#2pt}{#2pt}\hskip-#2pt
\rule{#2pt}{#1pt}}

\newcommand{\Yfund}{\raisebox{-.5pt}{\drawsquare{6.5}{0.4}}}
\newcommand{\Ysymm}{\Yfund\hskip-0.4pt%
                    \Yfund}
\def\symm{\Ysymm}
\def\bsymm{\overline{\Ysymm}}

\def\drawbox#1#2{\hrule height#2pt
        \hbox{\vrule width#2pt height#1pt \kern#1pt
              \vrule width#2pt}
              \hrule height#2pt}

\def\Fund#1#2{\vcenter{\vbox{\drawbox{#1}{#2}}}}
\def\Asym#1#2{\vcenter{\vbox{\drawbox{#1}{#2}
              \kern-#2pt       
              \drawbox{#1}{#2}}}}

\def\fund{\Fund{6.4}{0.3}}

\def\bfund{\overline{\fund}}

\newcommand{\be}{\begin{eqnarray}}
\newcommand{\ee}{\end{eqnarray}}

\begin{document}

\title{Composite Higgs from Higher Representations}
\author{Deog Ki Hong}
\email{dkhong@pusan.ac.kr} 
\affiliation{Department of Physics, Pusan National University,
Pusan 609-735, Korea}
\author{Stephen D.H.~Hsu}
\email{hsu@duende.uoregon.edu} 
\affiliation{Department of Physics, University of Oregon, Eugene
OR 97403-5203}
\author{Francesco Sannino}
 \email{francesco.sannino@nbi.dk}
\affiliation{NORDITA,  Blegdamsvej
  17, DK-2100 Copenhagen \O, Denmark }

\begin{abstract}
We investigate new models of dynamical electroweak symmetry
breaking resulting from the condensation of fermions in higher
representations of the technicolor group. These models lie close
to the conformal window, and are free from the flavor-changing
neutral current problem despite small numbers of flavors and
colors. Their contribution to the {\bf S} parameter is small and not
excluded by precision data. The Higgs itself can be light and
narrow.
\end{abstract}

\maketitle

\section{Introduction}

The simplest QCD inspired models of dynamical electroweak symmetry
breaking face a number of challenges: precision electroweak
constraints, suppression of flavor changing neutral currents,
generating the observed quark masses, and producing large enough
masses for any uneaten Goldstone bosons \cite{Hill:2002ap}. If the
strong dynamics at the electroweak scale is almost conformal
\cite{Holdom:1981rm,Yamawaki:1985zg,Appelquist:an} it is easier to
suppress flavor changing neutral currents on one hand, and to reduce
contributions to the {\bf S} parameter on the other
\cite{Peskin:1991sw,Sundrum:1991rf,Appelquist:1998xf,{Appelquist:1999dq}}.
We define near-conformal behavior by the smallness of the
$\beta$-function. When the $\beta$-function is small the gauge
coupling constant ``walks'' rather than runs over a large range of
energy.

In early models, a large number of fermions were needed to tune the
theory close to the conformal window \cite{Hill:2002ap}. Recently it
was shown that adding matter in two-index representations (either
symmetric or antisymmetric) of the technicolor group allows one to
reach the conformal window with only a small number of flavors and
colors \cite{Sannino:2004qp}. Indeed, $N_{T f}$ is constrained to be
$2 \leq N_{T f} \lesssim 4$ for any number of colors $N$. The
resulting contribution to {\bf S} is reduced significantly, even in
perturbation theory. Near the conformal window, {\bf S} is probably
reduced even more by non-perturbative dynamics, as we discuss below.
The models we consider are no more than 1-2 standard deviations from
the central value of precision data, according to our best
estimates. (Our models should not be confused with earlier work in
which matter in higher representations of QCD (``quixes'') were used
to break electroweak symmetry \cite{Marciano:zf}.)

Other positive features of our models are as follows (see below for
details). (1) The near-conformal behavior of the theory allows the
scale of fermion mass generation to be large, naturally suppressing
flavor changing neutral currents. (2) The mass of the composite
Higgs $m_H$ may be much lighter than expected in models of dynamical
symmetry breaking. Using a correspondence with supersymmetric
Yang-Mills theory which is exact at large $N$ and $N_{T f} = 1$, we
obtain a rough estimate of $m_H \sim 200-500$ GeV. (Note that in
this paper we only use supersymmetry as an analysis tool to obtain
Higgs mass estimates through a large-N correspondence - the models
themselves are not supersymmetric.) (3) In our favored models, which
have $N_{T f} = 2$, all Goldstone bosons resulting from symmetry
breaking are eaten by the electroweak gauge bosons. The favorable
features of $N_{T f} = 2$ might hint at the origin of the $SU_L(2)$
gauge symmetry.

This paper is organized as follows. First we specify several
models and give a table of some of their important properties.
Next we discuss, in turn, fermion mass generation, the {\bf S} parameter
and the Higgs mass and particle spectrum.

\section{Models}
The simplest technicolor model TC has $N_{T f}$ Dirac fermions in
the fundamental representation of $SU(N)$. These models, when
extended to accommodate the fermion masses through ETC interactions,
suffer from large flavor changing neutral currents. This problem is
alleviated, at least to the extent of accounting for masses up to
that of the b quark, if the number of flavors is sufficiently large
such that the theory is near conformal. This is estimated to happen
for $N_{T f} \sim 4 N$ \cite{Yamawaki:1985zg}, which implies a large
contribution to the {\bf S} parameter (at least in perturbative
estimates). We denote a generic, not near-conformal, technicolor
type model, with fermions in the fundamental representation, as
$TC(N,N_{T f})$. If it is near-conformal we use $WTC(N,N_{T f})$.

Near the conformal window \cite{Sundrum:1991rf,
{Appelquist:1998xf}} the {\bf S} parameter is reduced due to
nonperturbative corrections, but might still be too large if the
model has a large particle content. In addition, such models may
have a large number of unwanted pseudo Nambu-Goldstone bosons. By
choosing a higher dimension representation for the fermions one
can overcome these problems.

The simplest theories investigated in \cite{Sannino:2004qp} have
fermions in the two-index symmetric (S-type) or antisymmetric
(A-type) representation. In Table \ref{symmetric} we present the
generic S-type theory.
\begin{table}[h]
\begin{center}
\begin{minipage}{3in}
\begin{tabular}{c||ccccc }
 & $SU(N)$ & $SU_L(N_{T f})$& $SU_R(N_{T f})$&$U_V(1)$ & $U_A(1)$  \\
  \hline \hline \\
${q_{\{ij\}}}$& $\symm$ & $\fund$ & $1$ & $1$ & $1$  \\
 &&&\\
 $\widetilde{q}^{\{ij\}}$ & $\bsymm $ & $1$& $\bfund$& $-1$ & $1$ \\
 &&&\\
$G_{\mu}$ & {\rm Adj} & $0$&$0$ &$0$ & $0$    \\
\end{tabular}
\end{minipage}
\end{center}
\caption{The fermion sector in the case of the symmetric
representation. Here $q$ and $\widetilde{q}$ are Weyl fermions.}
\label{symmetric}
\end{table}
At infinite $N$ and with one Dirac flavor, the S and A type
theories become non-pertubatively equivalent to super Yang-Mills
\cite{Armoni:2004uu}. This property was used in making predictions
for QCD with one flavor \cite{{Armoni:2004uu},Sannino:2003xe} and
will be relevant also for our analysis. Theories of this type
theories emerge naturally in string theory via orientifold
projections.

The salient feature, found in \cite{Sannino:2004qp}, is that the
S-type theories are near conformal already at $N_{T f}=2$. This
should be contrasted with theories with fermions in the fundamental
representation for which the minimum number of flavors required to
reach the conformal window is eight. In the following $S(N,N_{T f})$
($A(N,N_{T f})$) represents an S(A)-type theory with $N$ colors and
$N_{T f}$ Dirac fermions.

The $N=3$ model with A-type fermions is just ordinary QCD with $N_f$
flavors and the maximum allowed number of flavors is $16$. For $N=2$
the antisymmetric representation goes over to the pure Yang-Mills
with a singlet fermion. For S-type models, asymptotic freedom is
lost already for three flavors when $N=2$ or $3$, while the upper
bound of $N_{T f}=5$ is reached for $N=20$ and does not change when
$N$ is further increased. A small number of flavors is generic to
the near conformal condition, which, as explained, is a favorable
feature for models of electroweak symmetry breaking.

In the theory with S-types, we therefore know that the number of
flavors must be smaller than $5$ for the theory to yield chiral
symmetry breaking. This takes into account that there is also a
conformal window of size $N_{T f}^c<N_{T f}<5$, with the critical
value $N_{T f}^c$ to be determined shortly. In \cite{Sannino:2004qp}
it has been shown that a theory with two S-types is very close to
the conformal window from $N=2$ up to a quite large $N$.

\begin{table}
\begin{tabular}{c||c|c|c}
  $G(N,N_{T f}/2)$ & {\bf S} & Higgs Mass & FCNC  \\
  \hline \hline &&&\\
  $TC(2,1)$ & ${1}/{3\pi}$ & $\sim\, 1\,{\rm TeV}$ & $\times$  \\
  $S(2,1)$ & ${1}/{2\pi} - \delta $ & $\sim 200 - 500~{\rm GeV}$  & \checkmark  \\
  $S(3,1)$ &  ${1}/{\pi} - \delta $&  $\sim 200 - 500~{\rm GeV}$  & \checkmark  \\
  $WTC(2,4)$ &  ${4}/{3\pi} - \delta$ & ? & \checkmark  \\
   $S(4,1)$ &  ${5}/{3\pi} - \delta $&  $\sim 200 - 500~{\rm GeV}$  & \checkmark  \\
  $A(4,4)$ & ${4}/{\pi} - \delta$ & $\sim 200 - 500~{\rm GeV}$& \checkmark  \\
  \hline
\end{tabular}
\caption{Properties of models discussed in the paper. $TC$ is
ordinary technicolor, $WTC$ walking technicolor, S models have
symmetric two-index matter, and $A$ models have antisymmetric
two-index matter. $N$ and $N_{T f}$ are colors and flavors.}
\end{table}

\section{Fermion masses and FCNC problem}
To generate fermion mass in technicolor models one needs
additional interactions, arising from extended technicolor (ETC),
which couple technifermions to  ordinary
fermions~\cite{Lane:1989ej}. However, the ETC interaction
typically leads to unacceptably large flavor-changing neutral
currents. This problem is less severe if the technifermion
bilinear, whose condensate breaks electroweak symmetry, has a
large anomalous dimension. This happens when the critical coupling
for triggering condensation is slightly larger than (but close to) the
infrared fixed point, $\alpha_c \ge \alpha_*$.\cite{footnote}

For $SU(N)$ gauge theories, the critical coupling is given in the
ladder approximation as $\alpha_c=\pi/(3C_2(R))$, where
$C_2(R)=(N+2)(N-1)/N$ for the second rank symmetric tensors. Since
the the coupling varies slowly up to a scale
$\Lambda_*\simeq\Lambda_{\rm
TC}\exp\left(\pi/\sqrt{\alpha_*/\alpha_c-1}\right)$, which is larger
than $300\Lambda_{\rm TC}$ for both $N=2,N_{T f}=2$ and $N=3,N_{T
f}=2$, we take the anomalous dimension $\gamma$ of the techni
bilinear to be close to unity \cite{Cohen:1988sq}, and
\begin{equation}
\left<\widetilde q\,q\right>_{\rm ETC} \simeq
\left({\Lambda_{\rm ETC}\over\Lambda_{\rm TC}}\right)\left<\widetilde
q q\right>_{\rm TC}.
\end{equation}
The enhancement of the condensate allows
reasonable masses for light quarks and leptons, even for
large ETC scales necessary to sufficiently suppress
flavor-changing neutral currents. However, to obtain the observed top mass,
we must rely on additional dynamics, as in so-called non-commuting
ETC models, where the ETC interaction does not commute with the
electroweak interaction~\cite{Chivukula:1994mn}.

If our goal is only to obtain an effective theory valid up to the
scale $\Lambda_{\rm ETC} \sim 10^3$ TeV, we need not explain the
origin of ETC operators (this is in the spirit of so-called
``little-Higgs'' models \cite{LH}). We leave an explanation of quark
and lepton masses for future work.

\section{Small S Parameter}
The models considered here produce smaller values of {\bf S} than
traditional technicolor models, because of the smaller particle
content and because of the near-conformal dynamics. The effect of
smaller number of particles can already be seen in Table 2, column
{\bf S}. The first number given in each entry is the perturbative
estimate, which is just $1/6\pi$ for each new electroweak doublet.
For S-type models the result is
\begin{equation}
{\bf S}_{\rm pert.} (S) = {1 \over 6 \pi} \cdot \frac{N(N+1)}{2}
\cdot \frac{N_{T f}}{2} \ ,
\end{equation}
while for $A$-type models it is
\begin{equation}
{\bf S}_{\rm pert.} (A) = {1 \over 6 \pi} \cdot \frac{N(N-1)}{2}
\cdot \frac{N_{T f}}{2} \ .
\end{equation}
However, near-conformal dynamics leads to a further reduction in
the {\bf S} parameter \cite{Sundrum:1991rf,Appelquist:1998xf}.
In the estimate of \cite{Sundrum:1991rf},
based
on the operator product expansion, the factor of ${1 \over 6 \pi}$
in the above equations is reduced to about $.04$, which is a
thirty percent reduction. For example, the best estimate for {\bf S}
in the S(3,1) model is about $.2$, which is within the $68\%$
confidence ellipse in the {\bf S-T} plane \cite{Hill:2002ap}.

\section{Light Higgs from Higher Representations}

In the analysis of QCD-like technicolor models one simply scales
up QCD phenomenology to obtain predictions at the electroweak
scale. The Higgs particle can then be mapped to the scalar chiral
partner of the Goldstone bosons which are eaten by the W and Z
gauge bosons. Naive scaling estimates yield a very heavy composite
Higgs with mass of the order a TeV: $m_H\sim 4 \pi F_{\pi}$, with
$F_{\pi}$ the electroweak scale.

There is, however, no guarantee that such estimates can be trusted
in WTC or near-conformal models. One cannot simply scale up QCD to
obtain useful nonpertubative information. In order to estimate the
Higgs mass we will use the observation made recently in
\cite{Armoni:2004uu} that {\em non}-supersymmetric Yang-Mills
theories with a Dirac fermion either in the two index symmetric or
antisymmetric representation of the gauge group are
nonperturbatively equivalent to supersymmetric Yang-Mills (SYM)
theory at large $N$, so that exact results established in SYM theory
should hold also in these ``orientifold'' theories. The orientifold
theories at finite $N$ were studied in \cite{Sannino:2003xe}, and
many of the discovered properties, such as almost parity doubling
and small vacuum energy density, are appealing properties for
dynamical breaking of the electroweak theory
\cite{Appelquist:1998xf}. We emphasize again that supersymmetry is
only used as a tool here to extract non-perturbative information
about our models, which are not themselves supersymmetric.

To estimate the Higgs mass for fermions in the S-type representation
of the gauge group we use the large $N$ limit while setting $N_{T
f}=1$ (since in our case of interest $N_{T f} = 2$ the results are
approximate, but probably roughly accurate). The Higgs particle is
then identified with the scalar fermion-antifermion state whose
pseudoscalar partner in ordinary QCD is the $\eta^{\prime}$. At
large $N$ this theory is mapped into super Yang-Mills using
\cite{Armoni:2004uu}. The low lying bosonic sector contains
precisely a scalar and a pseudoscalar meson. Due to the
supersymmetry correspondence the latter are expected to become
degenerate at infinite $N$. In the supersymmetric limit we can
relate the masses to the fermion condensate $\left< \widetilde{q}q
\right> \equiv \left< \widetilde{q}^{\{i,j \}} {q}_{\{i,j \}}
\right> $ \cite{Sannino:2003xe}:
\begin{eqnarray}
M=\frac{2\,\alpha}{3}\, \left[\frac{3 \left< \widetilde{q}q
\right> }{32\pi^2\,N} \right]^{\frac{1}{3}} =
\frac{2\hat{\alpha}}{3} \Lambda \ ,
\end{eqnarray}
with $\left< \widetilde{q}q \right> = 3N\Lambda^3$ and $\Lambda$
the one loop, large $N$, invariant scale of the theory:
\begin{eqnarray}
\Lambda^3 = \mu^3 \left(\frac{16\pi^2}{3Ng^2(\mu)}\right) \exp
\left[ \frac{-8\pi^2}{Ng^2(\mu^2)}\right] \ .
\end{eqnarray}
We have also defined:
\begin{eqnarray}
\hat{\alpha} = \alpha\, \left[\frac{9}{32\pi^2}
\right]^{\frac{1}{3}}\ .
\end{eqnarray}
The unknown numerical parameter $\hat{\alpha}$ is expected to be of
order one and is the coefficient of the K\"{a}hler term in the
Veneziano-Yankielowicz effective Lagrangian describing the lowest
composite chiral superfield. Taking for example $\hat{\alpha} \sim
1-3$ (see the discussion below) one would roughly deduce, at large
$N$ and for $N_{T f}=1$, a Higgs mass in the range:
\begin{eqnarray}
m_{H}=M\simeq 200-500~{\rm GeV} \ .
\end{eqnarray}
Here we have chosen $\Lambda =\Lambda_{TC}\sim 250~$GeV
\footnote{Strictly speaking, $F_{TC} = 250$ GeV. The relation
between $F_{TC}$ and $\Lambda_{TC}$ is given for two-index matter
(at large $N$) by $F_{TC} = c N \Lambda$. We took $c N$ of order
one; in QCD the large $N$ relation $F_{\pi} = c' \sqrt{N}
\Lambda_{QCD}$ implies $c'$ somewhat smaller than unity for $F_{\pi}
\sim 100$ MeV and $\Lambda_{QCD} \sim 300$ MeV, which is consistent
with $c N$ of order one.}. We expect $1/N$ corrections.
{}Fortunately these corrections were estimated, for $N_f=1$, in
\cite{Sannino:2003xe} and differ for theories of type S and A. {}For
the S-type models we have:
\begin{eqnarray}
\frac{m_H(S)}{M}=1 - \frac{4}{9N} + \frac{1}{8N}\frac{\langle
G_{\mu\nu}^a G^{a\mu\nu}\rangle}{\hat{\alpha}\,\Lambda^4} + {
O}(N^{-2}) \ ,
\end{eqnarray}
where $\langle G_{\mu\nu}^a G^{a\mu\nu}\rangle$ is the technigluon
condensate. Since $\langle G_{\mu\nu}^a G^{a\mu\nu}\rangle \sim
\Lambda^4$ and $\hat{\alpha}$ is order one the second term
dominates and further reduces the Higgs mass with respect to the
large $N$ limit. This should be compared to the $1/N$ corrections
for the A-type theory:
\begin{eqnarray}
\frac{m_H(A)}{M}=1 + \frac{4}{9N} + \frac{1}{8N}\frac{\langle
G_{\mu\nu}^a G^{a\mu\nu}\rangle}{\hat{\alpha}\,\Lambda^4} + {
O}(N^{-2}) \ ,
\end{eqnarray}
which indicate that the Higgs becomes heavier as we reduce the
number of colors. Since for $N=3$ the fermions, for type A
theories, are in the fundamental representation, our results are
qualitatively in agreement with the standard expectations that the
Higgs for theories with technifermions in the fundamental representation
is expected to be heavy.

To reassure ourselves that $\hat{\alpha}$ is, indeed, an order one
quantity we recall that the A-type theory with one flavor is mapped
into super Yang-Mills at large $N$. {}But, for $N=3$ the A-type
theory is QCD with one flavor, since the fundamental and two-index
antisymmetric representations are the same in $SU(3)$. This
observation was made long ago by Corrigan and Ramond
\cite{Corrigan:1979xf}. The $\eta^{\prime}$ state is the
pseudoscalar partner of the scalar fermion-antifermion state (i.e.
the Higgs) and its mass for one flavor can be simply estimated as
follows:
\begin{eqnarray}
m^2_{\eta^{\prime}}(N_f=1) = \frac{N_f}{3} m^2_{\eta^{\prime}} \ ,
\end{eqnarray}
where we used Witten and Veneziano's standard large $N$ and finite
$N_f$ scaling, with $N=3$. Comparing this mass with the
supersymmetric limit by identifying $M$ with $m_{\eta^{\prime}}$ we
estimate:
\begin{eqnarray}
\hat{\alpha} \sim \frac{\sqrt{3}}{2} \frac{m_{\eta^{\prime}}}{
\Lambda} \sim 3.2\, \frac{\sqrt{3}}{2} \sim 2.8\ .
\end{eqnarray}
Here $m_{\eta^{\prime}}=958~$MeV is the ordinary 3-flavor QCD mass
for the $\eta^{\prime}$ and $\Lambda\sim 300~$MeV is identified
with the characteristic QCD invariant scale. It is
encouraging that we obtained the
suggested order one result used earlier. Lattice
simulations should be able to improve the estimate for
the Higgs mass in our models.

Somewhat surprisingly we find a light scalar, presumably narrow.
Our calculations suggest that the favored S-type models naturally
produce light composite Higgs bosons.

\section{Conclusions}
In this letter we investigate a new class of models of dynamical
electroweak symmetry breaking with technifermions in higher
representations. These models lie close to the conformal window, and
are free from the flavor-changing neutral current problem despite
small numbers of flavors and colors. Their contribution to the {\bf
S} parameter is small and not excluded by precision data. Due to the
large $N$ equilvalence of our models to supersymmetric Yang-Mills
theory, we can make quantitative estimates of the mass of the Higgs.
It turns out to be surprisingly light (and perhaps narrow): $m_H
\sim 200-500~{\rm GeV}$.

The phenomenology of our models is quite distinct from QCD-like models
for two reasons:
the near-conformal behavior alters the dynamics, and the two-index
matter representations imply that color singlet interpolating operators
for bound states are very different from QCD.

\section*{Acknowledgements}
\noindent The authors thank H. Georgi and G. Veneziano for useful
comments, the Institute of Nuclear Theory at the University of
Washington for its hospitality and the Department of Energy for
partial support during the completion of this work. The work of
D.K.H. is supported by Korea Research Foundation Grant
(KRF-2003-041-C00073). The work of S.H. was supported in part under
DOE contract DE-FG06-85ER40224.



\begin{thebibliography}{9}
\bibitem{footnote} $\alpha$
never quite reaches the infrared fixed point, since fermions
decouple after chiral symmetry breaks. But $\alpha$ at the fermion
mass scale is very close to $\alpha_*$.
\end{thebibliography}
\end{document}